\documentclass[longnamesfirst,apjl]{emulateapj}
\submitted{Submitted to The Astrophysical Journal, Letters (Feb 8, 2004)}

\shorttitle{CONSEQUENCES OF GRAVITATIONAL RADIATION RECOIL}
\shortauthors{MERRITT ET AL.}

\newcommand{\lap}{\lesssim}
\newcommand{\gap}{\gtrsim}

\newcommand{\beq}{\begin{equation}}
\newcommand{\eeq}{\end{equation}}
\newcommand{\mh}{M_{\rm bh}}

\newcommand{\vk}{V_{\rm kick}}

\newcommand{\vrecoil}{V_{\rm eject}}
\newcommand{\ve}{V_{\rm esc}}

\newcommand{\me}{M_{\rm eff}}
\newcommand{\mb}{M_{\rm bound}}

\newenvironment{inlinefigure}{
\def\@captype{figure}
\noindent\begin{minipage}{0.95\linewidth}\begin{center}}
{\end{center}\end{minipage}\smallskip}

\begin{document}

\title{Consequences of Gravitational Radiation Recoil}

\author{David Merritt           \altaffilmark{1},
    Milo\v s Milosavljevi\'c    \altaffilmark{2},
    Marc Favata         \altaffilmark{3},
    Scott A.\ Hughes        \altaffilmark{4}, and
    Daniel E.\ Holz         \altaffilmark{5}}

\altaffiltext{1}{Department of Physics, Rochester Institute of Technology,
Rochester, NY 14623 drmsps@ad.rit.edu}

\altaffiltext{2}{Theoretical Astrophysics, California Institute of Technology,
Pasadena, CA 91125 milos@tapir.caltech.edu}

\altaffiltext{3}{Department of Astronomy, Cornell University, Ithaca,
NY 14853 favata@hewitt.astro.cornell.edu}

\altaffiltext{4}{Department of Physics, Massachusetts Institute of
Technology, Cambridge, MA 02139 sahughes@mit.edu}

\altaffiltext{5}{Center for Cosmological Physics, University of
Chicago, Chicago, IL 60637 deholz@cfcp.uchicago.edu}

\begin{abstract}

Coalescing binary black holes experience an impulsive kick
due to anisotropic emission of gravitational waves.
We discuss the dynamical consequences of the recoil accompanying massive
black hole mergers.
Recoil velocities are sufficient to eject most coalescing
black holes from dwarf galaxies and globular clusters,
which may explain the apparent absence of massive black holes
in these systems.
Ejection from giant elliptical galaxies would be rare,
but coalescing black holes are displaced from the center
and fall back on a time scale of order the half-mass
crossing time.
Displacement of the black holes transfers energy to
the stars in the nucleus and can convert a steep
density cusp into a core.
Radiation recoil calls into question models that grow supermassive
black holes from hierarchical mergers of stellar-mass precursors.
\end{abstract}

\keywords{black hole physics --- gravitation --- gravitational waves
--- galaxies: nuclei}

\section{Kick Amplitude}

In a companion paper (\citealt{Favata:04}; hereafter Paper I),
the amplitude of the recoil velocity experienced
by a binary black hole (BH) due to anisotropic emission
of gravitational radiation during coalescence
is computed.
Here we explore some of the consequences of the kicks
\citep{Redmount:89}:
the probability that BHs are ejected from galaxies and
the implications for BH growth;
the time scale for a kicked BH to return to the center
of a galaxy;
the effect of displacement on nuclear structure;
and other observational signatures of the kicks.
Unless otherwise indicated, notation is the same as in
Paper I.

For inspiral from a circular orbit,
the kick velocity is a function of the binary mass ratio
$q=m_1/m_2\le 1$,
the BH spins ${\tilde a}_1$ and ${\tilde a}_2$,
and the initial angle $\iota$
between the spin of the larger BH and the orbital angular
momentum of the binary.
Following Paper I, the spin of the smaller
BH is ignored.
Although Paper I only considers the cases $\iota=0$ and $\iota=180$, the
recoil for arbitrary inclination is likely to be bounded between these
extreme values.
Also, the detailed inclination dependence is
unimportant in comparison with the large uncertainty already present in
the contribution to the recoil from the final plunge and coalescence.
We will therefore assume that the restriction to
equatorial-prograde/retrograde orbits ($\tilde{a}_2=[-1,1]$)
considered in Paper I encompasses the characteristic range of recoil
velocities.

Figure 2b of Paper I shows upper- and lower-limit estimates
of the recoil velocity as a function of the effective spin parameter
$\tilde a$ for reduced mass ratio
$\eta=\mu/M=q/(1+q)^2=0.1$.
The {\em upper limit} for $\eta=0.1$ is well fit in the range
$-0.9\le\tilde a\le 0.8$
by the following fifth-order
polynomial:
\begin{eqnarray}
\label{eq:upper}
V_{\rm upper}&=&465\ {\rm km\ s}^{-1} {f(q)\over f_{\rm max}}(1-
0.281\tilde a -
 0.0361\tilde a^2 \nonumber \\
&-& 0.346\tilde a^3 - 0.374\tilde a^4
- 0.184\tilde a^5) .
\end{eqnarray}
Fitchett's (1983) scaling function $f(q)/f_{\rm max}$,
with $f(q)=q^2(1-q)/(1+q)^5$,
equals $0.433$ for $\eta=0.1$.
The {\em lower limit} curve of Paper I is well fit by
\begin{eqnarray}
\label{eq:lower}
V_{\rm lower}&=&54.4\ {\rm km\ s}^{-1} {f(q)\over f_{\rm max}}(1+
1.22\tilde a +
 1.04\tilde a^2 \nonumber \\
&+& 0.977\tilde a^3 - 0.201\tilde a^4
- 0.434\tilde a^5) .
\end{eqnarray}

We convert these expressions into estimates of the bounds on $\vk$ as follows.
First, as discussed in Paper I, there is an ambiguity in how
one translates the physical spin parameter ${\tilde a}_2$ of the
larger hole into the effective spin parameter ${\tilde a}$ of
equations (\ref{eq:upper}) and (\ref{eq:lower}).
Here we adopt the \citet{Damour:01} relation
$\tilde a=(1+3q/4)(1+q)^{-2}{\tilde a}_2$.
Second, Fitchett's scaling function assumes that
both bodies are non-spinning, and vanishes when $q=1$.
In fact, when $\tilde a\neq0$, significant recoil would occur even for $q=1$
due to spin-orbit coupling.
We can guess the approximate form of a new scaling function
by examining the spin-orbit corrections \citep{kidder} to Fitchett's
recoil formula.
For equatorial orbits, equation (4) of Paper I suggests that
$f(q)$ should be multiplied by the factor
$|1 + (7/29)\tilde{a}_2/(1-q)|/|1+ (7/29)\tilde{a}_2/(1-q')|$,
where $q'=0.127$ is the value used in defining $V_{\rm upper}$
and $V_{\rm lower}$ in equations (\ref{eq:upper}) and (\ref{eq:lower}).

Figure~\ref{fig:vplunge} plots upper and lower limits to $\vk$
as functions of ${\tilde a}_2$ and $q$.
The average over $\tilde{a}_2$ of the upper limit estimates
are $\sim(138, 444, 154)$ km s$^{-1}$
for $q=(0.1,0.4,0.8)$; Figure~\ref{fig:vplunge} suggests
a weak dependence on $\tilde{a}_2$.
Lower limit estimates are more strongly spin-dependent;
the averages over $\tilde{a}_2$
are $\sim(21.1,63.6,24.9)$ km s$^{-1}$ for the same values of $q$.
For moderately large spins ($\tilde{a}_2\gap 0.8$) and prograde
capture, the lower limit estimates exceed $100$ km s$^{-1}$
for $0.2\lap q\lap 0.6$.
In what follows, we will assume that
$\sim 500$ km s$^{-1}$ is an absolute upper limit to $\vk$.

\section{Escape}

When $V_{\rm kick}
\ge V_{\rm esc} \equiv \sqrt{2\phi({\bf r} = 0)}$,
with $\phi({\bf r})$ the
gravitational potential of the system (galaxy, dark matter halo)
hosting the BH,
the BH has enough kinetic energy
to escape.
Figure \ref{fig:escape} shows central escape velocities in four types of
stellar system that could contain merging BHs:
giant elliptical galaxies (E), dwarf ellipticals (dE),
dwarf spheroidals (dSph) and globular clusters (GC).
We fit the trend $\log\left(\ve/1\ {\rm km\ s}^{-1}\right)=\lambda-\beta M_V$
separately for each class of object.
dEs and GCs each separately establish a relation
$L\sim\ve^2$;
for GCs, this is compatible with the
relation found by \citet{Djorgovski:97}.
The E sample is consistent with the Faber-Jackson
(1976) relation.

\begin{inlinefigure}
\begin{center}
\resizebox{0.85\textwidth}{!}{\includegraphics{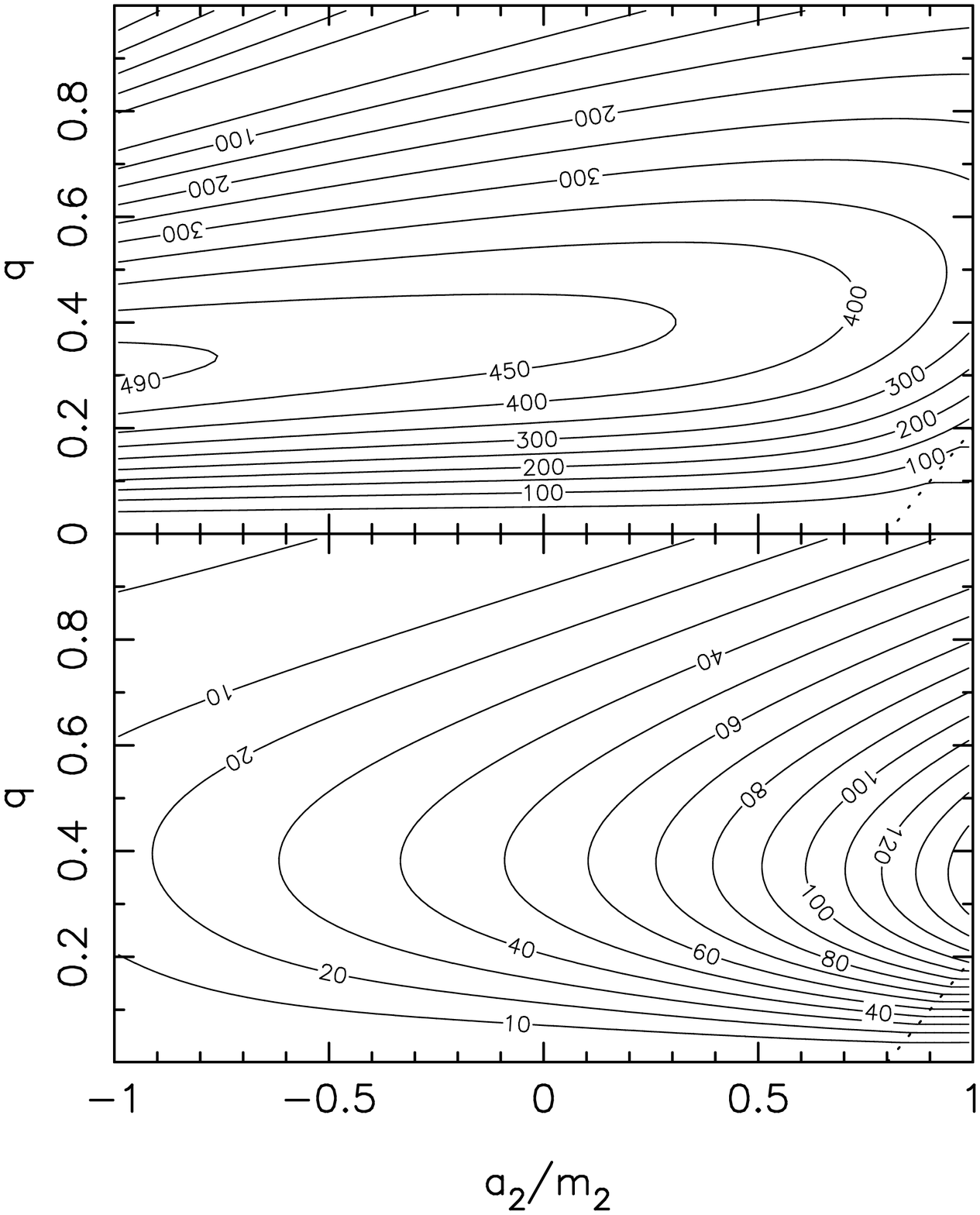}}
\end{center}
\figcaption{\label{fig:vplunge}
Upper limit (top) and lower limit (bottom) estimates of
$\vk$ as functions of mass ratio $q$ and spin of the larger
black hole ${\tilde a}_2$.
Units are km s$^{-1}$.
Values of ${\tilde a}_2$ and $q$ corresponding to $\tilde a>0.8$
lie in the region to the right of the dotted line.
Since equations (\ref{eq:upper}) and (\ref{eq:lower})
are not valid for $\tilde a>0.8$,
$\tilde a$ was replaced by $0.8$ in this
region.}
\end{inlinefigure}

The solid line in Figure \ref{fig:escape} shows escape velocities from the
dark matter (DM) halos associated with the luminous stellar systems.
To relate halo
properties to galaxy luminosities, we use the conditional luminosity
function $\Phi(L|M)dL$ from the concordance $\Lambda$CDM model M1 of
\citet{Yang:03}.
This function gives the number of galaxies in
the luminosity range $L\pm dL/2$ that reside inside a halo with virial
mass $M_{\rm vir}$.
The average luminosity $L_1$ of the brightest
(``central'') galaxy in the halo of mass $M_{\rm vir}$ is implicitly given
by the condition $\int_{L_1}^\infty\Phi(L|M_{\rm vir})dL=1$.
Inverting
this we obtain $M_{\rm vir}(L_1)$ and relate this mass to the escape
velocity via $\ve^2=2cg(c) GM_{\rm vir}/R_{\rm vir}$ where $R_{\rm
vir}$ is the virial radius of the halo, $c$ is the concentration of a halo
obeying the Navarro, Frenk \& White (1996; hereafter NFW) profile, and
$g(c)=[\ln(1+c)-c/(1+c)]^{-1}$
(e.g.,~\citealt{Lokas:01}).
Both $R_{\rm vir}$ and $c$ are functions of $M_{\rm vir}$ and the redshift $z$
(e.g.,~\citealt{Bryan:98,Bullock:01}).
At $z=0$ the average escape velocity is given by
$\ve=239\textrm{ km s}^{-1} (m_{11}/h)^{1/2}$,
where $M_{\rm vir}=10^{11}m_{11}M_\odot$ and $h$ is the
Hubble parameter, set to $0.7$ in Figure \ref{fig:escape}.
The upturn in
escape velocity for galaxies brighter than $M_V\sim-20$ is a consequence
of the increase in the occupation number of their host halos.
The dashed line in Figure \ref{fig:escape} shows the escape velocity from the
combined luminous + DM potential for the E galaxies, using the
scaling relation derived above to describe the luminous component.

Figure \ref{fig:escape} suggests that the consequences of the kicks are
strikingly different for the different classes of stellar system
that might host BHs.
Escape velocities from E galaxies
are dominated by the stellar contribution to the potential;
in the sample of Faber et al.~(1997), $\ve\gap
450\textrm{ km s}^{-1}$ even without accounting for DM.
This exceeds even the upper limits in Figure \ref{fig:vplunge}.
Hence, the kicks should almost never unbind BHs from E galaxies.
The tight correlations observed between BH mass and bulge
luminosity \citep{McLure:02,Erwin:03}\
and velocity dispersion \citep{FM:00,Gebhardt:00}
could probably not be maintained if escape occurred
with any significant frequency from luminous galaxies.

The existence of DM significantly affects the escape
probability from dE and dSph galaxies, implying kicks of
$\sim 300\textrm{ km s}^{-1}$ and $\sim 100\textrm{ km s}^{-1}$
respectively for escape.
In the absence of DM, these numbers would be
$\sim 100\textrm{ km s}^{-1}$ and $\sim 20\textrm{ km s}^{-1}$
respectively.
Hence, kicks of order 200 km s$^{-1}$ would unbind BHs
from dSph galaxies whether or not they contain DM,
while dE galaxies could retain their BHs if they are surrounded by
DM halos.

\begin{inlinefigure}
\begin{center}
\resizebox{\textwidth}{!}{\includegraphics{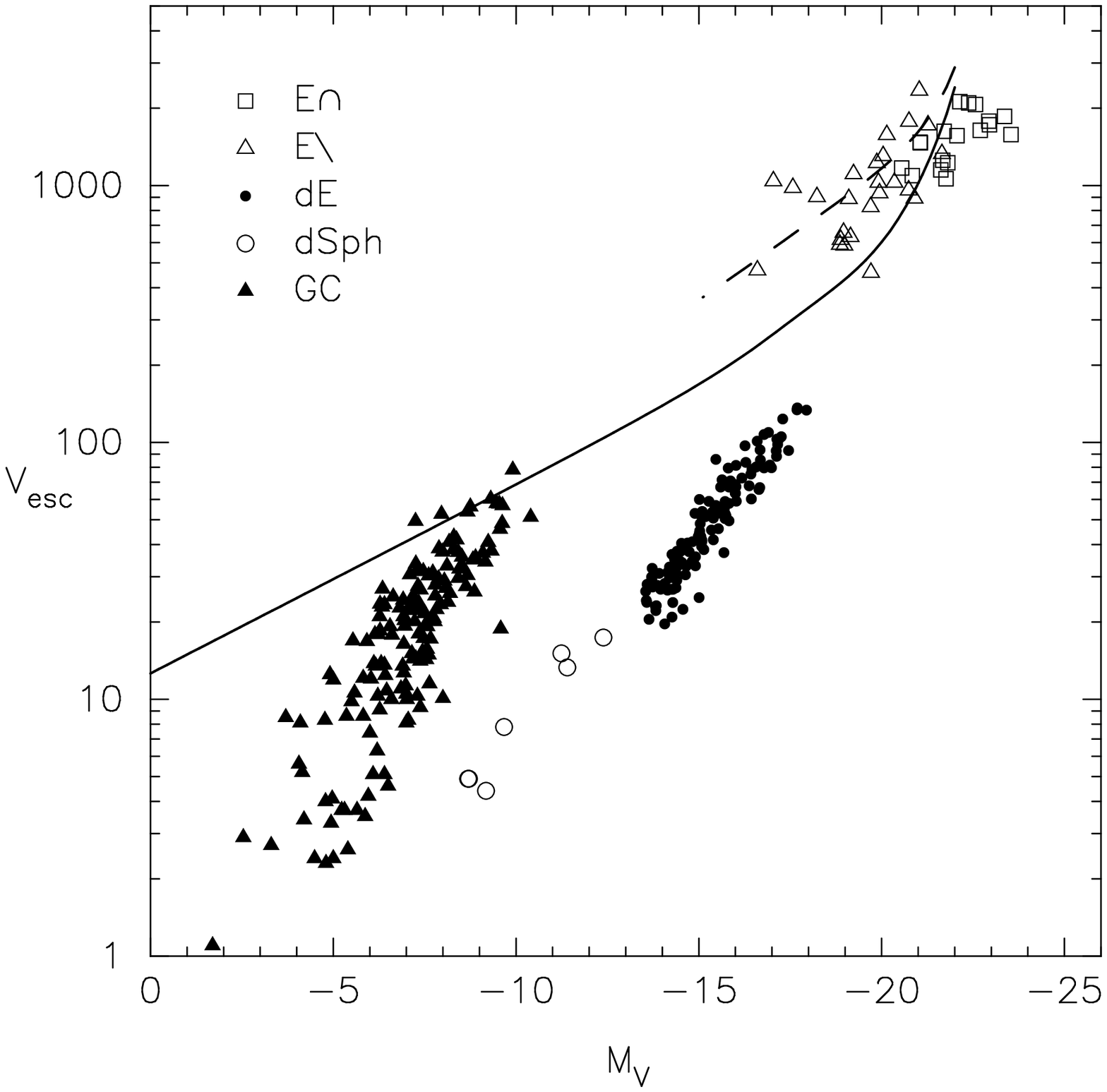}}
\end{center}
\figcaption{\label{fig:escape}
Central escape velocities in km s$^{-1}$ in four types of
stellar system that could harbor merging BHs.
E galaxy data are from \citet{Faber:97},
with separate symbols for core ($\square$)
and power-law ($\triangle$) galaxies.
dE data are from \citet{Binggeli:98}, with
mass-to-light ratios from \citet{Mateo:98}.
GC and dSph data are from the tabulation of \citet{Webbink:96}.
Solid line is the mean escape velocity from the DM
halos associated with the luminous matter.
Dashed line is the escape velocity from the combined
luminous + mean DM potentials for E galaxies.
}
\end{inlinefigure}

Evidence for intermediate-mass black holes
at the centers of galaxies fainter than
$M_V\approx -19$ is sketchy (e.g.,~\citealt{Marel:03}),
although there is indirect (non-dynamical) evidence
for BHs in faint Seyfert bulges \citep{FH:03}.
We note that the dense nuclei associated with
BHs in galaxies like M32 ($M_V\approx -19$)
become progressively less frequent at magnitudes
fainter than $M_V\approx -16$ and disappear entirely
below $M_V\approx -12$ \citep{Bergh:86}.
If the dense nuclei are associated with nuclear BHs
(e.g.,~\citealt{Peebles:72}), their absence
could signal loss of the BHs via ejection.
It is intriguing that these nuclei are sometimes
observed to be displaced far from the galaxy
center \citep{Binggeli:00}.
Figures \ref{fig:vplunge} and \ref{fig:escape}
imply that even kicks at the lower
limits of Paper I would almost always unbind BHs from GCs.

\section{Ejection in Hierarchical Merging Scenarios}
\label{sec:merging}

The kicks have serious implications for models in which massive BHs
grow from mergers of less massive seeds.  In
some of these models, the precursors are stellar- or intermediate-mass
black holes produced in the collapse of the first stars (Population III)
and the merging commenced in minihalos at redshifts as large as $\sim
20$ \citep{Madau:01,Volonteri:03,Islam:03}.
We evaluate the plausibility of such models in light of the
estimates of $\vk$ derived in Paper I.
Kicks from gravitational wave emission may compete with high-velocity
recoils \citep{Saslaw:74}
from (Newtonian) three-body interactions.
While the Newtonian recoil occurs only when
three BHs are present, which is contingent
on the galaxy merger rate and the BH binary orbital decay rate,
radiation recoil is present whenever BHs coalesce.

\begin{inlinefigure}
\begin{center}
\resizebox{\textwidth}{!}{\includegraphics{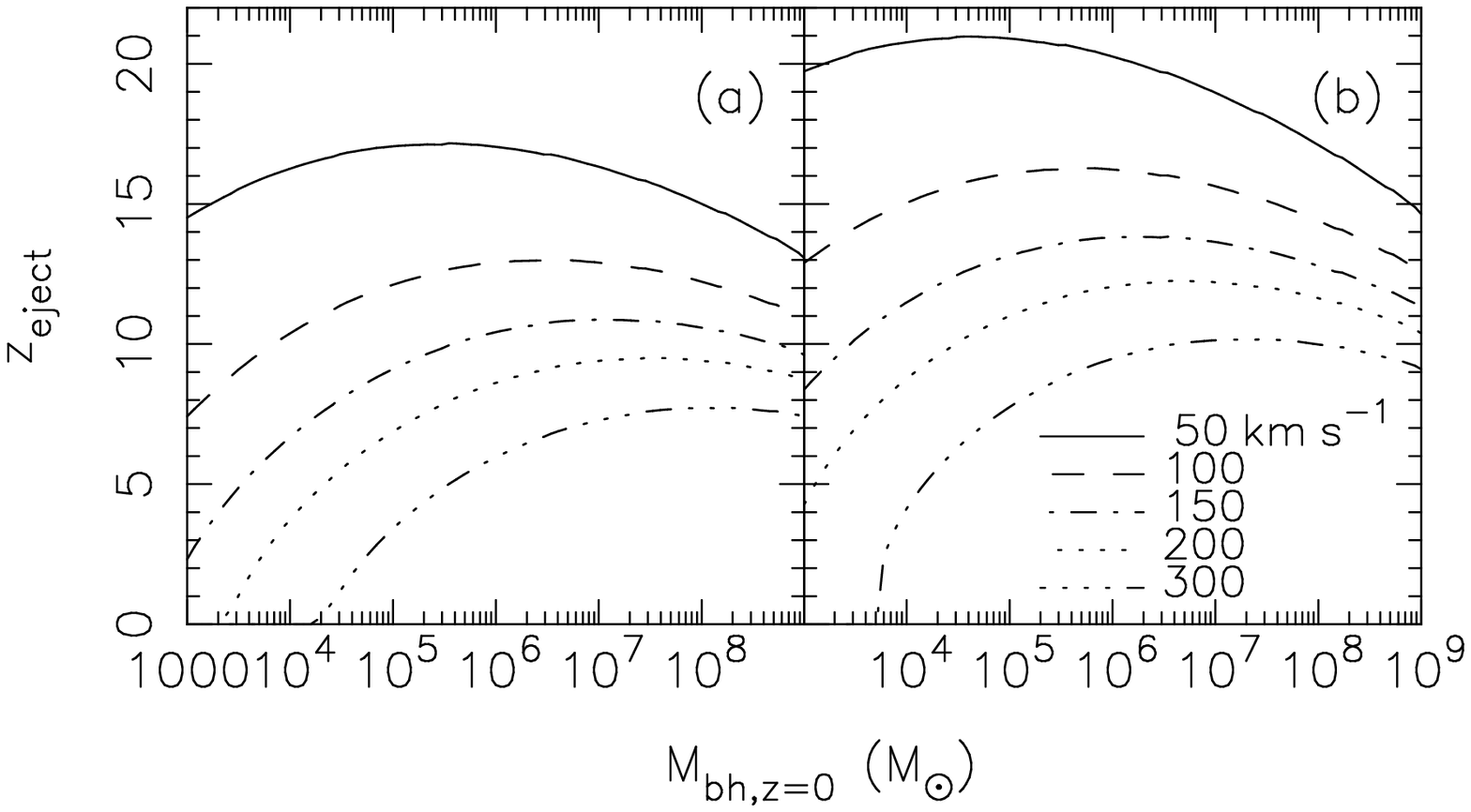}}
\end{center}
\figcaption{\label{fig:zeject}
The maximum redshift $z_{\rm eject}$ at which
(a) DM halos only,
and (b) DM halos and the central galaxies combined,
can confine
BHs as a function of the $z=0$ BH mass, for five values of the
kick velocity.  The depth of the galactic contribution to the potential
was calculated by identifying the velocity dispersion of the stellar spheroid
with the circular velocity of the halo \citep{Ferrarese:02}.}
\end{inlinefigure}

The confining effect of DM halos in a hierarchical universe
was smaller at higher redshifts when the average halo
mass was smaller.  We estimate the maximum redshift at which DM halos
can confine the progenitors of the present-day BHs.  \citet{Ferrarese:02}
derived a relation of the present-day BH mass $\mh(z=0)$ to the mass of
the host halo $M_{\rm vir}(z=0)$.
We use the \citet{Wyithe:02} form of the relation (their equation 11)
to obtain the host halo mass
and extrapolate the mass back in redshift via the
accretion history model \citep{Bullock:01} calibrated by \citet{Wechsler:02}
on a set of numerical simulations of DM clustering in a
$\Lambda$CDM universe.  The accretion trajectory
$M_{\rm vir}(z)\propto e^{-\alpha z}$, where $\alpha$ is itself a function
of the halo mass at $z=0$, can be interpreted as the mass of the most massive,
and thus the most easily confining parent halo at redshift $z$.
We can then calculate
the escape velocity $\ve(z)$ of the most massive progenitor
halo as a function of redshift.
Finally, we solve for $z_{\rm eject}$ such that $\vk=\ve(z_{\rm eject})$;
this is the maximum redshift at which the progenitors of the
present-day BHs could have started merging.
We also modelled the effect on $z_{\rm eject}$ of including the
potential due to a stellar component, idealized as an
isothermal sphere with core radius $r_h=2G\mh/\sigma^2$
and outer cutoff $10^3 r_h$; the 1D stellar velocity dispersion
$\sigma$ was related to halo circular velocity as in
Ferrarese (2002).

The results
for five representative
choices of $\vk$ are shown in Figure \ref{fig:zeject}.
For $\vk\sim150\textrm{ km s}^{-1}$, we find
$z_{\rm eject}< 11(14)$ over the entire range of $M_{bh}$;
the latter value is from the models that include a stellar
component.
For  $\vk\sim300\textrm{ km s}^{-1}$, the assembly of a
$10^8M_\odot$ BH must have started at $z\lesssim 8(10)$.
Models that grow supermassive BHs from mergers of seeds of
much lower mass at redshifts $z\gtrsim 10$
are thus disfavored due to the difficulty of retaining
the kicked BHs.
The effects of the kicks could be mitigated if early growth
of the BHs was dominated by accretion, or if BH mergers
were delayed until their halos had grown much more massive.

\section{Fallback Times}

A BH that has been kicked from the center
of a stellar system with a velocity less than $\ve$ falls back
and its orbit decays via dynamical friction against the stars and gas.
We define the fallback time $T_{\rm infall}$ as the time required
for a BH to return to a zero-velocity state after
being ejected.
The velocity with which the BH is ejected from the site of
the merger is $\vrecoil=(\mh/\me)\vk<\vk$; here $\me=\mh+\mb$ with
$\mb$ the mass in stars that remain bound to the BH after
it is kicked.
For recoil in a singular isothermal sphere nucleus $\rho\propto r^{-2}$,
$\me/\mh\approx (1.9,1.5,1.05,1.00)$ when
$\vk/\sigma = (0.5,1,2,3)$ where
$\sigma$ is the 1D stellar velocity dispersion;
$\mb/\mh\propto (\vk/\sigma)^{-4}$ for $\vk\gg\sigma$.

\begin{inlinefigure}
\begin{center}
\resizebox{\textwidth}{!}{\includegraphics{f4.ps}}
\end{center}
\figcaption{\label{fig:change}
Effect on the nuclear density profile of black hole ejection.
The initial galaxy model (black line) has a $\rho\sim r^{-1}$
density cusp.
(a) Impulsive removal of the black hole.
Tick marks show the radius of the black hole's sphere of influence
$r_h$ before ejection.
A core forms with radius $\sim 2r_h$.
(b) Ejection at velocities less than $\ve$.
The black hole has mass $0.003 M_{\rm gal}$;
the galaxy is initially spherical and the black hole's
orbit remains nearly radial as it decays via dynamical friction.
The arrow in panel marks $r_h$.
}
\end{inlinefigure}

We evaluated $T_{\rm infall}$ for BHs kicked from the
centers of Dehnen (1993)-law galaxies
for which
the central density obeys $\rho\propto r^{-\gamma}$.
Bright E galaxies have $0\lap\gamma\lap 1$
\citep{Gebhardt:96}, and cusps steeper than this
are likely to be softened by the binary BH prior to coalescence
\citep{Milos:01} and by the ejection itself (\S 5).
Given values for $\me$ and $\vrecoil$,
the fallback time in a spherical galaxy is given by the orbit-averaged
dynamical friction equation \citep{CK:78}.
For $\vrecoil/\ve\lap 0.6$, infall
times were found to be well approximated by
$T_{\rm infall}\approx T_{1/2}(\vrecoil/\ve)^{2.5(1+\gamma)}$
for $\me=0.001M_{\rm gal}$,
where the period $T_{1/2}$ of a circular orbit at the
galaxy's half-light radius is given in terms of the galaxy's
visual luminosity by
$T_{1/2}\approx 2\times 10^8\textrm{ yr}(L_V/10^{11}L_\odot)^{1/2}$
(\citealt{Valluri:98}).
Thus, return of a BH to a stationary
state requires of order a few times $10^8\textrm{ yr}$  or less
over a wide range of cusp slopes and galaxy luminosities
for $\vrecoil\lap \ve/2$.
As indicated in Figure~\ref{fig:escape},
this is the likely situation in the bright E galaxies.
Infall times are especially short for $\gamma\ge 1$,
since the BH experiences a strong impulsive frictional
force as it passes repeatedly through the dense center.
When $\vrecoil\lap\sigma$, the BH never moves far from its central
position and it carries much of the nucleus with it.  We carried
out $N$-body simulations of this regime and
found that return to zero velocity occurs in roughly
one orbital period
when $\vrecoil\lap\sigma$.
In fainter dE and dSph galaxies, ejection would more
often occur near $\ve$ and infall times could be
arbitrarily long, determined primarily by the mass
distribution at large radii.

In a nonspherical galaxy, an ejected BH
does not pass precisely through the dense center on each return,
delaying the infall.
To test the effect of non-spherical geometries on the infall time,
we carried out experiments in the triaxial generalizations of the
Dehnen models \citep{MF:96}.
Results were found to depend only weakly on the axis ratios
of the models.
Decay times in the triaxial geometry exhibit a spread in values
depending on the initial launch angle, bounded from below
by the decay time along the short axis.
We found a mean at every $\vrecoil/\ve$ that is $\sim 3-5$ times greater
than in a spherical galaxy with the same cusp slope.

\section{Observable Consequences of the Displacement}

Displacement of the BH also transfers energy to the nucleus
and lowers its density within a region of size $\sim r_h$,
the radius of the BH's sphere of influence
(defined here as the radius of a sphere containing a mass in
stars equal to twice that of the BH).
The simplest case to consider is $\vrecoil\gap\ve$;
the BH and its entrained mass depart the nucleus
on a time scale that is of order the crossing time at $r_h$ or less
and do not return.
The effect on the nucleus can be approximated by constructing
a steady-state model of a galaxy containing a central
point mass, then removing the point mass instantaneously and
allowing the remaining particles to relax to a new steady state.
Figure \ref{fig:change}a shows the results for three values of
$\me/M_{\rm gal}$.
Initial conditions consisted of $10^6$ particles
representing stars in a $\gamma=1$ Dehnen model.
We find that a core of roughly constant density forms
within a radius of $\sim 2r_h$.
Setting $\gamma=2$ (not shown) results in a core of size $\sim r_h$.
Figure \ref{fig:change}b shows the change in the nuclear density profile
for simulations with $\vrecoil<\ve$.
Significant changes in the central density require
$\vrecoil\gap 0.25\ve$.
We conclude that the recoil could affect the
observable structure of nuclei, since radii of
$\sim 2 r_h$ are resolved in many nearby galaxies
\citep{mf:01}.

The ``mass deficits'' seen at the centers of bright
galaxies \citep{Milos:02,Ravin:02} may be due to the
combined effects of slingshot ejection and BH displacement,
although we note that the large cores observed
in some bright galaxies could probably not be produced
by either mechanism \citep{Milos:02}.

The X-shaped radio sources associated with giant E galaxies
\citep{LP:92} are plausible sites of recent BH coalescence
\citep{ME:02}.
Displacement of the merged BHs from the galaxy center
prior to ignition of the ``active'' lobes
would imply a distortion of the X-morphology,
in the sense that the ``wings'' (the inactive lobes) would
be non-collinear near the center of the X.
Such distortions are in fact a common feature of the
X-sources \citep{Gopal:03},
although the linear scale of the distortions in some
of the X-sources (e.g. $\sim 10$ kpc in NGC 236;
\citealt{Murgia:01})
suggests that orbital motion of the merging galaxies
may be a more likely explanation \citep{Balcells:95}.

\smallskip
We thank D.~Axon, J.~Bullock, A.~Cooray,
A.~Klypin, M.~Santos, G.~Tormen, F.~van den Bosch,
and the anonymous referee for helpful comments.
DM is supported by NSF grant AST02-0631,
NASA grant NAG5-9046, and
grant HST-AR-09519.01-A from STScI.
MM is supported by a postdoctoral fellowship from the
Sherman Fairchild Foundation.
MF is partly supported by NSF grant PHY-0140209.
SAH is supported by NASA grant NAG5-12906 and by NSF
grant PHY-0244424.
DH is supported by NSF grant PHY-0114422.

\end{document}